\documentclass[12pt]{article}
\usepackage{amsmath,amssymb,amsthm}
\usepackage{txfonts,bm,pifont}
\usepackage{graphicx}
\usepackage[pdfencoding=auto,bookmarks=true,bookmarksnumbered=true,colorlinks=true]{hyperref}
\usepackage[svgnames]{xcolor}
\AtBeginDvi{}
\theoremstyle{plain}
\newcommand{\R}{\mathbb{R}}

\usepackage[tight]{subfigure}

\makeatletter
\providecommand*{\diff}{\@ifnextchar^{\DIfF}{\DIfF^{}}}
\def\DIfF^#1{\mathop{\mathrm{\mathstrut d}}\nolimits^{#1}\gobblespace}
\def\gobblespace{\futurelet\diffarg\opspace}
\def\opspace{\let\DiffSpace\!\ifx\diffarg(\let\DiffSpace\relax\else\ifx\diffarg[\let\DiffSpace\relax\else\ifx\diffarg\{\let\DiffSpace\relax\fi\fi\fi\DiffSpace}

\providecommand*{\deriv}[3][]{\frac{\diff^{#1}#2}{\diff #3^{#1}}}

\providecommand* {\R}[1][] {\mathbb{R}^{#1}}

\providecommand* {\norm}[1]{\left\lVert#1\right\rVert}
\providecommand* {\ord}[1]{\mathcal{O}(#1)}

\DeclareMathOperator*{\argmin}{arg\,min}
\DeclareMathOperator{\sn}{sn}
\DeclareMathOperator{\cn}{cn}

\makeatletter
\@addtoreset{equation}{section} 

\makeatother
\begin{document}
\begin{center}
\begin{Large}
Fairing of Discrete Planar Curves by Discrete Euler's Elasticae\\[5mm]
\end{Large}
\begin{normalsize}
Sebasti\'an El\'ias {\sc Graiff Zurita}\\[1mm]
Graduate School of Mathematics, Kyushu University\\
744 Motooka, Fukuoka 819-0395, Japan\\
e-mail: s-graiff@math.kyushu-u.ac.jp\\[2mm]
Kenji {\sc Kajiwara}\\[1mm]
Institute of Mathematics for Industry, Kyushu University\\
744 Motooka, Fukuoka 819-0395, Japan\\
e-mail: kaji@imi.kyushu-u.ac.jp\\[2mm]
\end{normalsize}
\end{center}
\begin{abstract}
After characterizing the integrable discrete analogue of the Euler's elastica, we focus our
attention on the problem of approximating a given discrete planar curve by an appropriate discrete
Euler's elastica.  We carry out the fairing process via a
$L^2\!$-distance minimization to avoid the numerical instabilities.  The
optimization problem is solved via a gradient-driven optimization method (IPOPT).  This problem is
non-convex and the result strongly depends on the initial guess, so that we use a
discrete analogue of the algorithm provided by Brander et al., which gives an initial guess to the
optimization method.
\end{abstract}
% keywords (Mandatory field): 3-5 Keywords
%\keywords{Euler's elastica, integrability, discrete curve, discrete differential geometry, fairing}
\section{Introduction}
Recently, Brander et al. \cite{BRAN2017} have constructed an algorithm to fair a given planar curve
segment by an Euler's elastica, motivated mainly by the development of the robotic hot-blade cutting
technology. In this work, we construct a discrete version of that algorithm, which would be used to
fair a given discrete planar curve by an integrable discrete analogue of Euler's elastica.
The definition in this paper is a well-known expression appearing frequently in different forms in the literature on the discrete integrable systems \cite{Bobenko-Suris:1999,Hoffmann-Kuts:2004,SOGO2006}.
Actually, the Euler's elastica is known to be equivalent to the simple pendulum, and to have a close relationship with the Lagrange top, as shown in \cite{Bobenko-Suris:1999}, for example.

\section{Euler's elastica}
Let $\gamma(s)\in\R[2]$ be an arc length parameterized planar curve ($s$: arc length), and define the
tangent and normal vectors as $T(s) = \gamma'(s)$ and $N(s) = R_{\pi\!/2}T(s)$, respectively, where
$R_{\pi\!/2}$ is a $\pi\!/2$ rotation matrix and ${\text{ }}'=\diff{}/\diff{s}$. By definition,
$\norm{T(s)} = 1$, so the tangent vector can be parameterized as
\[ T(s) = \begin{pmatrix}
    \cos\psi(s)\\
    \sin\psi(s)
  \end{pmatrix},\]
where $\psi(s)$ is the angle function, namely, the angle of the tangent vector measured from the
horizontal axis, so that $\kappa(s) = \psi'(s)$ is the curvature.  With this notation, a plain curve
satisfying
\begin{equation} \label{eq:psi}
  \psi'' + \mu\sin(\psi -\phi) = 0,
\end{equation}
or equivalently
\begin{equation} \label{eq:kappa}
  \kappa'' + \frac{1}{2}\kappa^3 - \lambda\kappa = 0,
\end{equation}
where $\mu > 0$, and $\phi, \lambda \in \R$ are constant parameters, is called an Euler's elastica.
These curves can also be characterized as critical points of the elastic energy
\begin{equation} \label{eq:functional.E}
  E[\gamma] = \int_0^L\frac{1}{2}\psi'^2(s)\diff{s} = \int_0^L \frac{1}{2}\kappa^2(s)\diff{s},
\end{equation}
when the total length $L$ is fixed. Furthermore, \eqref{eq:psi} and \eqref{eq:kappa} are reductions
of the sine-Gordon equation and the modified KdV equation, respectively, where the latter describes
the integrable deformation of planar curves.

\section{Discrete Euler's elastica}
Let $\gamma_n \in\R[2]$ ($n = 0, \dots, N-1$) be a discrete planar curve with constant segment length $h = \norm{\gamma_{n+1}-\gamma_n}$, and
define the tangent vector as $T_n = (\gamma_{n+1} - \gamma_n)/h$, which we will parameterize as in
\cite{KAJI2012}:
\begin{eqnarray} 
&  T_n = \begin{pmatrix}
    \cos\psi_n\\
    \sin\psi_n
  \end{pmatrix},\label{eq:discrete.tangent}\\
&  \psi_n = \dfrac{\theta_{n} + \theta_{n + 1}}{2}.\label{eqn:d-angle}
\end{eqnarray}
Here, $\psi_n$ is the discrete angle function and 
$\theta_n$ plays the role of a potential function.
Finally, noticing that $\psi_{n}-\psi_{n-1} = (\theta_{n+1}-\theta_{n-1})/2$ is the 
angle between $T_{n-1}$ and $T_n$, we introduce the discrete curvature $\kappa_n$ by
\begin{equation}
  \kappa_n = \frac{2}{h} \tan\left(\frac{\theta_{n+1} - \theta_{n-1}}{4}\right).
\end{equation}
With these notations, a discrete planar curve satisfying
\begin{equation} \label{eq:def}
  \tan\left(\frac{\theta_{n+1} + \theta_{n-1}}{4}\right) = \frac{1 - h^2\mu/4}{1 + h^2\mu/4} \tan\left(\frac{\theta_n}{2}\right),
\end{equation}
or equivalently
\begin{equation} \label{eq:def.sin}
   \sin\!\left(\frac{\theta_{n+1}-2\theta_n+\theta_{n-1}}{4}\right)
+\frac{h^2\mu}{4} \sin\!\left(\frac{\theta_{n+1}+2\theta_n+\theta_{n-1}}{4}\right)\!=\!0, 
\end{equation}
for some constant $\mu > 0$, will be referred to as an integrable discrete analogue of Euler's elastica (or simply, as a discrete elastica); see, for example, equation (9) in \cite{SOGO2006}.  These curves can also be characterized as critical
points of the functional
\begin{equation} \label{eq:functional.S}
  \scalebox{0.9}{$\displaystyle
    S[\theta_*]\!=\!\sum_{n = 0}^{N-2}h\!\left[\frac{\sin^2\left(\frac{\theta_{n+1} - \theta_n}{4}\right)}{h^2/8}\!+\!\mu \cos\left(\frac{\theta_{n+1} + \theta_n}{2}\right)\right]\!, $}
\end{equation}
when the endpoints $\theta_0$ and $\theta_{N-1}$ are fixed.  Furthermore, as well as in the
continuous case, \eqref{eq:def} can be seen as a reduction of the discrete sine-Gordon equation,
\begin{equation}\label{eqn:d-sG}
\sin\left(\frac{\theta_{n+1}^{m+1} - \theta_n^{m+1} - \theta_{n+1}^m + \theta_n^m}{4}\right) 
= \frac{a_n}{b_m} \sin\left(\frac{\theta_{n+1}^{m+1} + \theta_n^{m+1} + \theta_{n+1}^m + \theta_n^m}{4}\right),  
\end{equation}
where $a_n$ an $b_m$ are arbitrary functions in the indicated variables.
Note that \eqref{eqn:d-sG} can be rewritten as 
\begin{equation}
    \tan\left(\frac{\theta_{n+1}^m + \theta_{n}^{m+1}}{4}\right)= \frac{1 - \varepsilon_n^m}{1 + \varepsilon_n^m}  \tan\left(\frac{\theta_{n+1}^{m+1} + \theta_{n}^m}{4}\right),
\end{equation}
with $\varepsilon_n^m = a_n/b_m$. Then, after regarding $\varepsilon = \varepsilon_n^m$ as a
constant, and applying the change of variables $k = n - m$ and $l = n + m$, we get
\begin{equation}
    \tan\left(\frac{\theta_{k+1}^{l} + \theta_{k-1}^{l}}{4}\right) = \frac{1 - \varepsilon}{1 + \varepsilon} \tan\left(\frac{\theta_{k}^{l+1} + \theta_{k}^{l-1}}{4}\right),
\end{equation}
which reduces to \eqref{eq:def}, if we neglect the $l$--dependence and replace $\varepsilon =
h^2\mu/4$.  Lastly, it can be seen that the difference equation \eqref{eq:def} possesses the conserved
quantity
\begin{equation} \label{eq:conserved.A}
    \cos\left(\frac{\theta_{n+1} - \theta_n}{2}\right) 
+ \frac{h^2\mu}{4} \cos\left(\frac{\theta_{n+1} + \theta_{n}}{2}\right) = A,
\end{equation}
where $A\in\R$ is a constant fixed by the initial condition.
\par\bigskip
\noindent{\bf Remark 1.}
Note that, in this context, the discrete angle function \eqref{eqn:d-angle}
has a simple expression, but it has imposed some difficulties when trying to find an {\em energy}
expression for the discrete curve $\gamma_n$.  In the simplest case, let us try to minimize the
functional \eqref{eq:functional.S} with $\gamma_n$ as the independent variable.  First, we need to
establish a one-to-one relation between $\theta_n$ and $\gamma_n$, which can be done by fixing, for
example, the end point $\theta_{N-1} = 0$.
This allow us to write
  \begin{equation}
    \theta_n = \sum_{m = n}^{N-2} (-1)^{m-n}\,2\psi_m.
  \end{equation}
So, after solving the variational problem $\delta S[\gamma_*] = 0$ we recover the equation
\eqref{eq:def} for the inner points ($n = 1, \dots, N-2$), but we also
need to satisfy
\begin{equation}
    \sum_{n=0}^{N-2}(-1)^n \, 2\delta\psi_n = 0, \label{eq.con1} 
\end{equation}
or
\begin{equation}
\tan\left(\frac{\theta_1}{2}\right) = \frac{1 - h^2\mu/4}{1 + h^2\mu/4} 
\tan\left(\frac{\theta_0}{2}\right). \label{eq.con2} 
\end{equation}
The first condition \eqref{eq.con1} gives a non-local constraint for the perturbation, so we might
want to satisfy the second condition \eqref{eq.con2}.  If we choose so, we will impose a boundary
condition in the definition of the discrete elastica, which is undesired; among other reasons, the
continuous limit will only reach a subset of the Euler's elastica family. Because of this, we have
chosen to describe a discrete elastica by the functional \eqref{eq:functional.S}, defined over the
potential function $\theta_n$.
\par\bigskip

%-- REMARK 2 -- 
\noindent{\bf Remark 2.}
Let $\widehat{S}[\theta_*]$ be the functional presented by Sogo in \cite{SOGO2006},
\begin{equation} \label{eq:functional.Sogo}
\scalebox{0.9}{$\displaystyle
 \widehat{S}[\theta_*]\!=\!\sum_{n = 0}^{N-1}\!
\left[\sin^2\!\left(\frac{\theta_{n+1}\!\!-\!\!\theta_n}{4}\right) 
-\varepsilon \sin^2\!\left(\frac{\theta_{n+1} + \theta_n}{4}\right)\!\right].
$}
\end{equation}
So, we can see that, after replacing $\varepsilon = h^2\mu /4$, the functionals
\eqref{eq:functional.S} and \eqref{eq:functional.Sogo} are related as
\begin{equation}
 S[\theta_*] = \frac{8}{h} \left( \widehat{S}[\theta_*]
\Big|_{\varepsilon = h^2\mu/4}\right) + \frac{h\mu N(N-1)}{2}.
\end{equation}
Thus, both share the same critical points, and so they describe the same discrete curves.  Moreover,
we believe that this approach shows a more transparent connection between the continuous and
discrete case; it can be seen that the functional \eqref{eq:functional.S} is a discrete
approximation, of order one, of the continuous variational problem, i.e.
\begin{equation}
 \int_0^h \left[\frac{1}{2} \left(\psi'(s)\right)^2 +\mu \cos\psi(s)\right] \diff{s} 
 = S[(\psi(0), \psi(h))] + \ord{h^2}.
\end{equation}
Actually, the summand of the discrete functional \eqref{eq:functional.S} yields by putting $s=nh$
\begin{align}
& h\left[\frac{\sin^2\left(\frac{\psi(s+h) - \psi(s)}{4}\right)}{h^2/8}
+\mu \cos\left(\frac{\psi(s+h) + \psi(s)}{2}\right)\right] & \nonumber\\
&= h\left[\frac{\left(\frac{h\psi'(s)}{4}\right)^2}{h^2/8}+\mu \cos\psi(s)\right] + \ord{h^2}\nonumber\\
&= h\left[\frac{1}{2} \left(\psi'(s)\right)^2 +\mu \cos\psi(s)\right] + \ord{h^2}.
\end{align}

\subsection{General solution and characteristic parameters}
Here we will just present the result without its derivation. We refer to \cite{SOGO2006}
for a detailed explanation.

It can be shown that the general solution of \eqref{eq:def} is
\begin{equation} \label{eq:gen.disc}
  \sin \frac{\theta_n}{2} = k \sn(q + zh n,k),
\end{equation}
along with
\begin{equation}
  \mu = \frac{4}{h^2}\left(\frac{1 - \cn(zh,k)}{1 + \cn(zh,k)}\right).
\end{equation}
where $k\geq 0$, $q, z\in\R$ are parameters that can be determined from the initial condition
$\gamma_0$, $\gamma_1$, and $\gamma_2$. Then, since we are going to implement the fairing process for
arbitrary discrete curves, we incorporate the freedom of rotation via the parameter $\phi\in\R$. So,
by applying the transformation $\theta_n \rightarrow \theta_n - \phi$, the general solution changes
to
\begin{equation} \label{eq:disc}
  \sin \frac{\theta_n - \phi}{2} = k \sn(q + zh n,k).
\end{equation}
Finally, to reconstruct the discrete elastica we also need the starting point $\gamma_0 ={}^{t}(x_0, y_0)$.
So, the discrete elastica can be obtained via
$\gamma_n = \gamma_{n-1} + h T_{n-1}$ ($n = 1,\dots, N-1$), and parameterized by seven parameters:
\begin{equation} \label{eq:param}
  p = (x_0, y_0, h, \phi, z, q, k).
\end{equation}

\section{Fairing process}
\subsection{Outline}
In order to approximate a given discrete curve by a discrete elastica we must choose a suitable
criteria.  One possibility is to minimize the functional \eqref{eq:functional.S}; however, having
$N$ free parameters ($\theta_0, \theta_1, \dots, \theta_{N-1}$),  makes the fairing process numerically unstable.  Furthermore, if we minimize it with
respect to the discrete curve $\gamma_n$ instead of the potential function
$\theta_n$, from Remark 1 we know that the constraints imposed on the starting points will
restrict the domain of all compatible elasticae.  Because of this, we have decided to do the fairing
process via a $L^2$-distance minimization.  Namely, we seek to find a set of parameters $p^*$ that
minimize
\begin{equation} \label{eq:min}
  \mathcal{L}(p) := \sum_{n = 0}^{N-1} \frac{1}{2}\norm{\zeta_n(p) - \gamma_n}^2,
\end{equation}
where $\zeta_n(p)$ is the discrete elastica generated by $p$.

The optimization problem,
\begin{equation} \label{eq:opt}
  p^* = \argmin_{p} \left\{\sum_{n = 0}^{N-1} \frac{1}{2}\norm{\zeta_n(p) - \gamma_n}^2 \right\},
\end{equation}
is non-convex and the result strongly depends on the initial guess. So, we have decided to implement
a discrete analogue of the algorithm provided in \cite{BRAN2017}, which provides an initial guess to
the IPOPT method.

So, from here and after applying a discrete analogue of the algorithm shown in \cite{BRAN2017}, we
have observed that it is possible to recover the seven control parameters \eqref{eq:param} that
characterize a given discrete elastica segment, in a numerically stable manner.

\begin{figure}[h!]
\includegraphics[width=5.2cm]{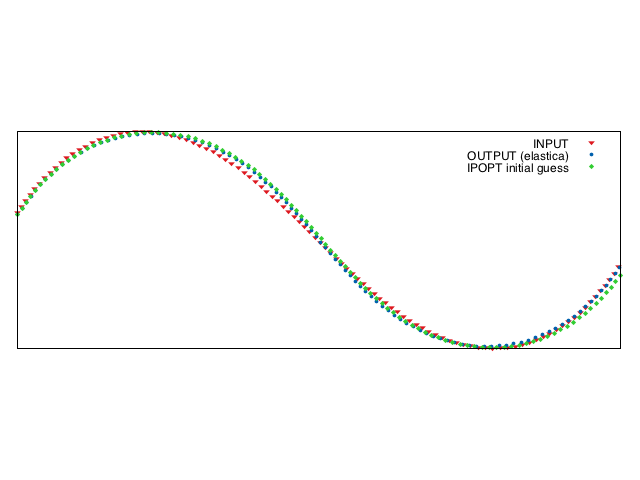}
\includegraphics[width=5.2cm]{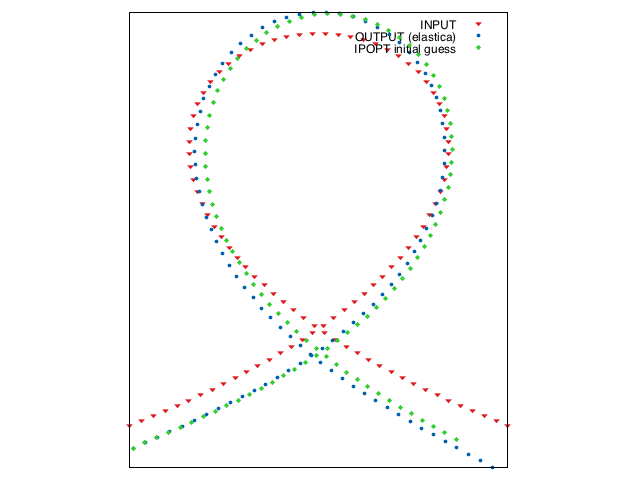}
\includegraphics[width=5.2cm]{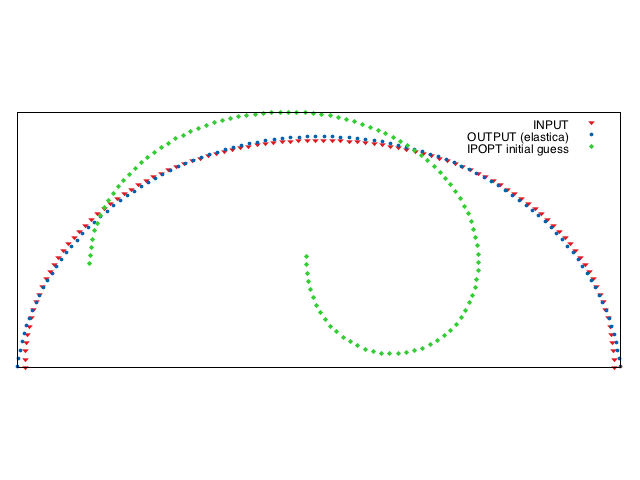}
  \caption{Examples of discrete curves approximated by Euler's elasticae, 
displayed for three different values of the parameter $k$. (a) left: $k < 0.9089...$ (lemniscate-like curve), (b) center: $0.9089... < k < 1$, (c) right: $k > 1$. Red: given discrete curve, blue: output found with IPOPT optimization, green: initial guess.}
  \label{fig:examples}
\end{figure}

Some examples are shown in Figure \ref{fig:examples}, where the optimization problem was solved via
a gradient-driven optimization method (IPOPT \cite{IPOPT}).

\subsection{Some remarks on the algorithm}
The algorithm shown in \cite{BRAN2017} is anchored on the following: in a continuous context, define
$u(s)\in\R$ as the projection of $\gamma(s)$ onto the line spanned by $(\sin\phi, -\cos\phi)$, and
$\psi_u(s)$ as the angle between the tangent $T(s)$ and that same line, i.e.:
\begin{align}
  u(s) &= (\sin\phi, -\cos\phi)\cdot\gamma(s), \label{eq:cont.1}\\
  \cos \psi_u(s) &= (\sin\phi, -\cos\phi)\cdot T(s), \label{eq:cont.2}\\
  \sin \psi_u(s) &= (\cos\phi, \sin\phi)\cdot T(s). \label{eq:cont.3}
\end{align}
Thus, using the fact that $u' = \cos\psi_u$ and $\psi_u'=\psi'$, we have
\begin{equation} \label{eq:eq1}
  \deriv{}{u}\sin\psi_u = \frac{1}{u'}\,\deriv{}{s}\sin\psi_u = \psi' = \kappa.
\end{equation}
Then, from the differential equation \eqref{eq:psi} and the former expressions, it follows that
\begin{align}
  \kappa(s) &= \mu u(s) + \alpha, \text{ and } \label{eq:affine.cont}\\
  \sin\psi_u &= \frac{1}{2}\mu u^2 + \alpha u + \beta, \label{eq:alg.2}
\end{align}
where $\alpha, \beta \in\R$ are constants of integration. From this point, all the parameters that
characterize the smooth Euler's elastica can be re-obtained, after solving several quadratic
minimization sub-problems and using the explicit expressions for the curvature $\kappa(s)$, the
angle function $\psi(s)$, and the curve itself $\gamma(s)$.

For our case, define $u_n$ and $\psi_{u_n}$ in the same way as the smooth case,
\begin{align}
  u_n &=  (\sin\phi, -\cos\phi)\cdot\gamma_n, \label{eq:def.un}\\
  \sin \psi_{u_n} &= (\cos\phi, \sin\phi)\cdot T_n.
\end{align}
We can show that (see Section \ref{sec:affine}) the affine relation \eqref{eq:affine.cont} between $\kappa(s)$ and
$u(s)$ holds also in the discrete framework, i.e.:
\begin{equation}
  \kappa_n = \frac{1}{A} \mu u_n + \alpha, \label{eq:affine.disc}
\end{equation}
where $A$ is the conserved quantity \eqref{eq:conserved.A}, which can be easily checked that it goes
to 1 in the limit $h \to 0$.
Finally, for the discrete version of the equation \eqref{eq:alg.2} we have chosen to consider its
approximation up to order one,
\begin{equation}
  \sin \psi_{u_n} = \frac{1}{2} \mu u_n^2 + \alpha u_n + \beta + \ord{h^2},
\end{equation}
and similarly for the rest of the quadratic minimization sub-problems that follows.

\section{Affine relation for the discrete curvature} \label{sec:affine}
In this last section we will see how the affine relation for the continuous curvature \eqref{eq:affine.cont} has an equivalent expression \eqref{eq:affine.disc} in the discrete framework.

First, and in order to simplify the notation, let us consider the following quantities:
\begin{align}
  \varphi_n &= \frac{\theta_{n+1} - \theta_n}{2},\\
  K_n &= \frac{\theta_{n+1} - \theta_{n-1}}{2} = \varphi_n + \varphi_{n-1} = \psi_n - \psi_{n-1}.                    
\end{align}
With this notation, the discrete Euler's elastica equation \eqref{eq:def} reads as
\begin{equation} \label{eq:def2}
  \sin\left(\frac{\varphi_n - \varphi_{n-1}}{2}\right) = -\frac{h^2\mu}{4} \sin\left(\frac{\psi_n + \psi_{n-1}}{2}\right),
\end{equation}
and the discrete curvature as
\begin{equation} \label{eq:kappa.def}
  \kappa_n = \frac{2}{h} \tan\left(\frac{K_n}{2}\right).
\end{equation}
We can reorder the terms in \eqref{eq:def2}, in two different ways, to obtain
\begin{align}
  \sin\left(\frac{K_n}{2} - \varphi_{n-1}\right) &= -\frac{h^2\mu}{4} \sin\left(\frac{K_n}{2}+\psi_{n-1}\right),\\
  \sin\left(\varphi_n - \frac{K_n}{2}\right) &= -\frac{h^2\mu}{4} \sin\left(\psi_n - \frac{K_n}{2}\right).
\end{align}
Then, after using the addition formula for the sine function, and the expression for the conserved quantity $A$, equation \eqref{eq:conserved.A}, the previous two equations can be rearranged as
\begin{align}
  A\tan\left(\frac{K_{n}}{2}\right) &= \sin\varphi_{n-1} - \frac{h^2\mu}{4}  \sin\psi_{n-1}, \label{eq:tan1}\\
  A\tan\left(\frac{K_{n}}{2}\right) &= \sin\varphi_{n} + \frac{h^2\mu}{4}  \sin\psi_{n},    \label{eq:tan2}.
\end{align}
So, if we compute the difference between \eqref{eq:tan1}, with $n \mapsto n+1$, and \eqref{eq:tan2}, we get
\begin{equation}
  A\left(\tan\left(\frac{K_{n+1}}{2}\right) - \tan\left(\frac{K_{n}}{2}\right)\right) = -\frac{h^2\mu}{2} \sin \psi_n,
\end{equation}
which can be rewritten in terms of the discrete curvature $\kappa_n$ given in \eqref{eq:kappa.def} as
\begin{equation}
  \kappa_{n+1} - \kappa_n = -\frac{h\mu}{A} \sin\psi_n.
\end{equation}

Finally, introducing the rotation angle $\phi$ to replace $\theta_n \mapsto \theta_n - \phi$, 
the last equation is modified as
\begin{equation}
  \kappa_{n+1} - \kappa_n = -\frac{1}{A} \mu h \sin(\psi_n - \phi).
\end{equation}
By using the definition \eqref{eq:def.un} of $u_n$ we get
\begin{equation}
  \kappa_{n+1} - \kappa_n = \frac{1}{A} \mu\,(u_{n+1} - u_n),
\end{equation}
which implies that
\begin{equation}
  \kappa_n - \frac{1}{A} \mu u_n = \alpha,
\end{equation}
with $\alpha\in\R$ constant.
\par\bigskip
%%%%%%%%%%%%%%%%%%%%%%%%%%%%%%%%%%%%%%%%%%%%%
\noindent\textbf{Acknowledgements.}\\[2mm]
The authors are grateful to Professor David Brander for his comments and discussions. They also
thank Professor Toshitomo Suzuki for his interest in this work in view of the application to
architecture. They also acknowledge the support of Joint Use Program of Institute of Mathematics for
Industry, Kyushu University: 2016 Short-term Joint Research ``Differential Geometry and Discrete
Differential Geometry for Industrial Design''(September 2016) and 2017 Short-term Joint Research
``New Development of Discrete Differential Geometry: From Industrial Design to Architecture''
(September 2017). This work is supported by JSPS KAKENHI Grant No. JP16H03941 and 
JST CREST Grant No. JPMJCR1911, Japan.


\begin{thebibliography}{99}
\bibitem{BRAN2017}
D. Brander, J. Gravesen and T. B. N\o rbjerg, Approximation 
by planar elastic curves, Adv. Comput. Math., {\bf 43} (2017), 25--43.
\bibitem{Bobenko-Suris:1999}
A.I. Bobenko and Y.B. Suris, Discrete time Lagrangian mechanics on Lie groups, with an application to
the Lagrange top, Commun. Math. Phys., {\bf 204} (1999), 147--188.
\bibitem{Hoffmann-Kuts:2004}
T. Hoffmann and N. Kutz, 
Discrete curves in $\Bbb C\rm P^1$ and the Toda lattice,
Stud. in Appl. Math. {\bf 113} (2004), 31--55.
\bibitem{SOGO2006}
 K. Sogo, Variational discretization of Euler's elastica problem, J. Phys. Soc. Jpn., {\bf 75} (2006), 064007.
\bibitem{KAJI2012}
  J. Inoguchi, K. Kajiwara, N. Matsuura and Y. Ohta, Motion and B\"acklund transformations of discrete planar curves, Kyushu J. Math., {\bf 66} (2012), 303--324.
\bibitem{IPOPT}
  A. W\"achter and L.T. Biegler, On the implementation of an interior-point filter line-search algorithm for large-scale nonlinear programming, Math. Program., Ser. A {\bf 106} (2006), 25--57.
\end{thebibliography}
\end{document}